\documentclass[aps,10pt,a4paper,jcp,twocolumn]{revtex4-1}

\usepackage{amsmath}
\usepackage{lineno}
\usepackage{subfigure}
\usepackage{hyperref}
\usepackage{natbib}
\usepackage{algorithm}
\usepackage{mathtools}
\usepackage{nicefrac}
\usepackage[noend]{algpseudocode}
\usepackage{graphicx}
\usepackage[english]{babel} 

\newcommand{\appropto}{\mathrel{\vcenter{
  \offinterlineskip\halign{\hfil$##$\cr
    \propto\cr\noalign{\kern2pt}\sim\cr\noalign{\kern-2pt}}}}}

\newcommand{\dir}{.}

\begin{document}


\title{
Simulating Copolymeric Nanoparticle Assembly in the Co-solvent Method: 
How Mixing Rates Control Final Particle Sizes and Morphologies
} 

\author{Simon Ke\ss ler}
\affiliation{Institut f\"ur Physik, Johannes Gutenberg-Universit\"at Mainz,
   D 55099 Mainz}

\author{Klaus Drese}
\affiliation{ Fakult\"at f\"ur Angewandte Naturwissenschaften, 
  Hochschule Coburg, 96450 Coburg}

\author{Friederike Schmid}
\affiliation{Institut f\"ur Physik, Johannes Gutenberg-Universit\"at Mainz,
   D 55099 Mainz}

\begin{abstract}

The self-assembly of copolymeric vesicles and micelles in micromixers is
studied by External Potential Dynamics (EPD) simulations -- a dynamic density
functional approach that explicitly accounts for the polymer architecture both
at the level of thermodynamics and dynamics.  Specifically, we focus on the
co-solvent method, where nanoparticle precipitation is triggered by mixing a
poor co-solvent into a homogeneous copolymer solution in a micromixer.
Experimentally, it has been reported that the flow rate in the micromixers
influences the size of the resulting particles as well as their morphology: At
small flow rates, vesicles dominate; with increasing flow rate, more and more
micelles form, and the size of the particles decreases. Our simulation model is
based on the assumption that the flow rate mainly sets the rate of mixing of
solvent and co-solvent. The simulations reproduce the experimental observations
at an almost quantitative level and provide insight into the underlying
physical mechanisms: First, they confirm an earlier conjecture according to
which the size control takes place in the earliest stage of the particle
self-assembly, during the spinodal decomposition of polymers and solvent.
Second, they reveal a crossover between different morphological regimes as a
function of mixing rate. Hence they demonstrate that varying the mixing rate in
a co-solvent setup is an effective way to control two key properties of drug
delivery systems, their mean size and their morphology. 

\end{abstract}


\maketitle


\section{INTRODUCTION}

Nanoparticles are molecule aggregates with spatial extensions on scales of
$\sim$ 10 to 100 nm.  They have attracted growing interest during the last
decades due to improved technologies for visualization and manipulation of
nanoscale structures that have revealed their great potential for a wide range
of applications. Depending on the chemical composition of the nanoparticles, these applications
include mesoscopic models for atomic systems \cite{Herlach.2010,
Smallenburg.2011}, optoelectronic devices \cite{Bucaro.2009}, nanoreactors,
models for biological cells \cite{Lensen.2008, Discher.2002}, and most
prominently, transport vehicles for medication, i.e. drug delivery systems
\cite{Zhang.2008, Gindy.2009, Vartholomeos.2011}. 

Most commercially available drug delivery systems are liposomes
\cite{Allen.2004, Bleul.2014}, which are vesicular particles made of
am\-phi\-phi\-lic lipids. Biocompatible amphiphilic (diblock) copolymers
represent promising alternatives to the lipids, since they are more stable and
can be synthesized and modified more easily \cite{Discher.1999, Thiermann.2014,
Thiermann.2012}. In analogy to their lipid counterparts, vesicles built from
amphiphilic di\-block-co\-po\-ly\-mers are often called polymersomes
\cite{Discher.1999}. In the context of drug delivery systems one key property of nanoparticles is their size, since it not only determines their
loading capacity, but also the composition of their protein corona in blood \cite{Tenzer.2011},
which in turn affects retention times in the circulatory system. In addition,
the nanoparticle size plays a critical role in passive targeting of tumors
based on the Enhanced Permeability and Retention effect \cite{Maeda.2000}. 

A method to manufacture nanoparticle populations with a specific mean size is
the co-solvent method, also known as flash nanoprecipitation \cite{Zhang.2012}.
In this method, an initial solution containing the molecular constituents of
the nanoparticles (e.g., the copolymers) in good solvent is mixed with a
selective or bad solvent. The mixing eventually triggers the precipitation of
the constituents and the aggregation to particles.  One way to tune the sizes
of the particles is to vary the rate of solvent mixing \cite{Mueller.2009,
Thiermann.2012, Thiermann.2014, Nikoubashman.2016}. In experiments, solvent
mixing is often implemented by continuous flow mixing devices, called passive
micromixers \cite{Mueller.2009, Thiermann.2012, Thiermann.2014,
Nikoubashman.2016}. Mixing rates in passive micromixers increase with
increasing flow rates $v$ \cite{Falk.2010, Drese.2004, Schonfeld.2004.2}, and
in turn, particle sizes are found to decrease with increasing $v$.  Thiermann
et al.\ \cite{Thiermann.2012, Thiermann.2014} have recently carried out an
extensive study on the relation between flow rates and particle size in such a
micromixer setup.  The constituents of their nanoparticles were amphiphilic
diblock-copolymers with a hydrophobic polybutadiene block and a hydrophilic
polyethylene-oxide block, and they used tetrahydrofuran as good solvent and
water as selective co-solvent.  They found that variations of $v$ for symmetric
flow conditions in passive micromixers enable a reproducible control over the
mean size $R$ of a particle population with relatively low polydispersity.  The
experimental data for $R$ could be described reasonably  well by scaling laws
$R\propto v^{\alpha}$, where the exponent $\alpha$ differs from measurement to
measurement, but scatters around a mean value of $\alpha=-0.158$ with a
standard deviation $\sigma_{\alpha}=0.058$ \cite{simon_diss}.

In a recent publication we proposed an explanation for the experimentally
observed nanoparticle size dependence on flow rates or equivalently, mixing
times \cite{meinArtikel1}. We combined a simple Cahn-Hilliard equation with a
Flory-Huggins-de Gennes free energy functional for {\em homo}polymers and
implemented solvent mixing by a time dependent interaction parameter. The
simulation results for the size of \textit{homopolymer} aggregates at the
(relatively well-defined) crossover between spinodal decomposition and
subsequent coarsening were in good semiquantitative agreement with the
experimentally determined sizes for stabilized \textit{co\-po\-ly\-mer}
particles from Thiermann et al \cite{Thiermann.2012,Thiermann.2014} (apart from
a factor of two).  This lead us to hypothesize that particle sizes in the
co-solvent method are determined during the very early stages of phase
separation. The description also provided an explanation for the typical
scaling laws observed in experiments and predicted an analytical expression, $R
\sim v^{-\alpha}$ with the exponent $\alpha = 1/6$, which is in good agreement
with the experimentally observed exponents. According to this theory, the
particle sizes at different $v$ result from a competition of the repulsion
between solvent molecules and monomers of the solvent-phobic block with the interfacial tension of diffuse
interfaces in the very early stages of phase separation, during spinodal
decomposition.  With respect to the entire particle growth process the
description in terms of the Cahn-Hilliard model is, however, incomplete.  The
Flory-Huggins-de Gennes free energy functional for homopolymers can only be
applied in the early segregation regime, it does not describe the internal
reorganization of copolymers inside the particles at later stages.  In
particular, it does not include a mechanism that physically stabilizes
particles of finite size, and cannot be used to describe particles with
more complex morphologies as can be observed in copolymeric systems.

Amphiphilic molecules may form particles of different morphologies in solution,
including spherical micelles or vesicles \cite{Discher.2002}. Apart from the
size, particle morphology is another key property of drug delivery systems
because it affects the loading possibilities.  Spherical micelles consist of a
hydrophobic core surrounded by a shell of hydrophilic blocks and they can only
be loaded with hydrophobic substances.  Vesicles allow both hydrophilic and
hydrophobic loading, which makes them appealing canditates for multifunctional
drug delivery systems: Hydrophilic substances can be enclosed in their solvent
containing core, while the hydrophobic part of the bilayer shell can be loaded
with hydrophobic substances. These loading possibilities are important because
some therapeutic substances are hydrophilic and others hydrophobic. An example
for a hydrophilic substance is the toxic anti-cancer drug camptothecin
\cite{Wall.1966}.  Hydrophilic materials are, for instance, the anti-cancer
drug doxorubicin \cite{Torchilin.2005} or the dye pho\-lo\-xine B, which can be
used to trace particles in {\em in vitro} cell binding studies or studies on
(hydrophilic) loading efficiencies \cite{WMueller.2011}. Although vesicles are
more versatile when it comes to loading possibilities, micelles have the
advantage that, due to their smaller size compared to vesicles, they may 
enable ways of cellular uptake that bypass the drug efflux mechanism of cancer
cells in order to treat multiresistant cancer \cite{Bleul.2014, Kabanov.2002}.

The experiments by Thiermann et al. \cite{Thiermann.2012, Thiermann.2014}
clearly show that the flow rates in micromixers also affect the morphologies of
the resulting particles. At small flow rates, vesicles dominate, while at
larger flow rates, micelles become more frequent. Motivated by these
observations, we here present a numerical study of copolymer aggregation in a
co-solvent setup for different mixing rates, using a dynamic density functional
approach that explicitly accounts for the molecular architecture of copolymers.
To this end, we implement time dependent interaction parameters into the
established 'External Potential Dynamics'' (EPD) model \cite{Maurits.1997b},
which has been successfully used to study spontaneous self-assembly of
amphiphilic diblock-copolymers to nano\-particles \cite{He.2006, He.2008}.  We
focus specifically on the effect of mixing rates on particle morphologies, and
how the existence of different particle morphologies affects the typical
scaling behavior $R\propto v^{\alpha}$. 

The article is organized as follows. In section \ref{sec: theoretical_model} we
present the theoretical model. In section \ref{sec: simulation_setup} we
specify the input parameters used in the current article. The results and the
discussion is presented in section \ref{sec: results_and_discussion}, and the
summary is given in \ref{sec: summary_and_outlook}. In the \ref{sec:
numerical_integration_scheme}, we discuss technical issues and describe a novel
integration scheme which was used to perform the simulations.

\section{THEORETICAL MODEL}
\label{sec: theoretical_model}

We consider a solution of an amphiphilic AB-diblock copolymer P and a single
solvent S in a volume $V$ at temperature $T$. To describe the phase separation
dynamics we apply the EPD model \cite{Maurits.1997b, Muller.2005, He.2006},
which is based on the free energy functional of the popular Self
Consistent Field (SCF) Theory for polymers \cite{Edwards.1965, Schmid.1998}:
\begin{equation}
\begin{aligned}
\begin{split}
\frac{\beta \:F}{n} =& - f_S \ln\left(\frac{Q_S}{Vf_S}\right) - \frac{f_P}{N}\ln\left(\frac{Q_P N}{V\, f_P}\right) + \frac{1}{V}\int_{V}\Big[ -\omega_A\phi_A 
\\ &   -\omega_B\phi_B -\omega_S\phi_S+\chi_{AB}\phi_A\phi_B+\chi_{AS}\phi_A\phi_S 
\\ &  +\chi_{BS}\phi_B\phi_S   + \frac{\kappa_H}{2}\left(\phi_A+\phi_B+\phi_S -1 \right)^2 \Big]\, d\vec{r} .
\end{split}
\end{aligned}
\label{eq7: SCF_freie_Energie}
\end{equation}
Here $f_P$ and $f_S$ are the mean polymer and solvent volume fractions, $N$ is
the number of monomers per polymer chain, $\kappa_H$ a mean compressive modulus
of the solution \cite{Schmid.1998, Helfand.1975}, $\chi_{ij}$ the
Flory-Huggins interaction parameter between species $i$ and $j$, and $\beta =
1/k_B T$ the Boltzmann factor. The fields $\omega_i$ for $i=A$, $B$, $S$ are
potentials in units of $\beta^{-1} = k_B T$ that act on the respective monomer
species $i$, and $\phi_i=\nicefrac{\rho_i}{\rho_0}$ with
$\rho_0=\nicefrac{n}{V}$ are normalized number densities, where $n=n_S+N n_P$
is the total number of monomers and solvent molecules in the system. $n_S$ is the number of solvent
molecules and $n_P$ the number of polymer chains. $Q_P$ is the partition
function of a polymer chain subject to $\omega_A$ and $\omega_B$, while $Q_S$
is the partition function of a single solvent molecule exposed to $\omega_S$.
These single chain partition functions are given by
\begin{equation}
Q_P=\int_Vg(\vec{r},1)\, d\vec{r} \text{ and } Q_S =\int_V e^{-\omega_S(\vec{r})}\, d\vec{r}.
\end{equation}
$g(\vec{r},s)$ is the end-segment distribution function and describes the
probability that one end of a polymer chain segment of length $s$ is located at
position $\vec{r}$. It obeys the inhomogeneous diffusion equation 
\begin{equation}
\begin{aligned}
\frac{\partial g}{\partial s}(\vec{r},s)=\Delta g(\vec{r},s) - N \omega g(\vec{r},s) 
\text{ with }
g(\vec{r},0)=1
\end{aligned}
\label{eq7: diffusionsgl_g}
\end{equation}
\cite{Helfand.1972, Fredrickson.2006}, where omega is defined by
\begin{equation}
\omega = \begin{dcases*}
\omega_A & $0 < s < c_A$ \\
\omega_B & $c_A < s < 1$\\
\end{dcases*}.
\label{eq7: omega_g}
\end{equation}
$\vec{r}$ denotes the position in units of the polymer's radius of
gyration $R_g$, $s\in[0,1]$ represents the position along a polymer chain in
units of $N$, and $c_A$ specifies the fraction of $A$-monomers in the diblock
copolymer. If the polymer chain contains $N_A$ monomers of type $A$, one has
$c_A=\nicefrac{N_A}{N}$. Since the potential fields in the SCF Theory mimic
mean interactions between different species, they depend on the densities.
Introducing a second distribution function $g'$, which also solves equation
(\ref{eq7: diffusionsgl_g}) with 
\begin{equation}
\omega = \begin{dcases*}
\omega_B & $0 < s < c_B$\\
\omega_A & $c_B < s < 1$\\
\end{dcases*}
\label{eq7: omega_g'}
\end{equation}
and $c_B=\nicefrac{N_B}{N}=1-c_A$, the relation between the potential fields
$\omega_i$ and the densities $\phi_i$ from the SCF Theory can be cast into the
form
\begin{equation}
\phi_A(\vec{r})[\omega_A]=\frac{V f_P}{Q_P}\int_0^{c_A} g(\vec{r},s) g'(\vec{r},1-s)\, ds,
\label{eq7: bestimmung_phiA}
\end{equation}
\begin{equation}
\phi_B(\vec{r})[\omega_B]= \frac{V f_P}{Q_P}\int_{c_A}^{1} g(\vec{r},s) g'(\vec{r},1-s)\, ds,
\label{eq7: bestimmung_phiB}
\end{equation}
and
\begin{equation}
\phi_S(\vec{r})[\omega_S]=\frac{V f_S}{Q_S}e^{-\omega_S(\vec{r})}.
\label{eq7: bestimmung_phiS}
\end{equation}
In the EPD formalism the dynamical equations for the potential fields are 
given by
\begin{equation}
\begin{split}
\frac{\partial \omega_i}{\partial t}(\vec{r},t) 
= - D_i \Delta\left(\mu_i(\vec{r},t) + \eta_i(\vec{r},t) \right) 
\text{ with } i = A,B,S,
\\ 
\mu_i(\vec{r})=\frac{1}{\rho_0}\frac{\delta \beta F}{\delta\phi_i(\vec{r})}
\text{, and } D_i = 
\begin{dcases*}
D_P=\frac{D_0}{N} & $i=A,B$ \\
D_S=D_0 & $i=S$ \\
\end{dcases*},
\end{split}
\label{eq7: entwicklungsgleichung_omega}
\end{equation}
where $\eta_i$ is a random noise and $\frac{\delta \beta F}{\delta \phi_i}$ is
the variational derivative of $\beta F$ with respect to $\phi_i$. Equation
\ref{eq7: entwicklungsgleichung_omega} is equivalent to the dynamical master
equation for monomer densitiy fields derived by Kawasaki and Sekimoto
\cite{Kawasaki.1987} with a non-local kinetic coupling in a copolymer solution.
The EPD formalism dramatically reduces the computational cost of the non-local
coupling model and was first introduced by Maurits and Fraaije
\cite{Maurits.1997b}. Taking into account the density-field relations from the
SCF Theory and that $Q_P=Q_P[\omega_A,\omega_B]$ as well as $Q_S=Q_S[\omega]$,
the chemical potentials $\mu_i$ can be calculated as
\begin{equation}
\mu_A = \chi_{AB}\phi_B + \chi_{AS}\phi_S + \kappa_H \left(\phi_A + \phi_B +\phi_S - 1\right) -\omega_A,
\label{eq7: variationsableitung_F_phi_A_EPD}
\end{equation}
\begin{equation}
\mu_B= \chi_{AB} \phi_A + \chi_{BS}\phi_S + \kappa_H (\phi_A +\phi_B+\phi_S -1)-\omega_B,
\end{equation}
\begin{equation}
\mu_S = \chi_{AS} \phi_A + \chi_{BS} \phi_B +\kappa_H(\phi_A+\phi_B+\phi_S-1)-\omega_S.
\label{eq7: variationsableitung_F_phi_S_EPD}
\end{equation}

Equations \ref{eq7: diffusionsgl_g}, \ref{eq7: bestimmung_phiA} -- \ref{eq7:
bestimmung_phiS}, and \ref{eq7: entwicklungsgleichung_omega} constitute a
SCF-EPD model that has been used to successfully study self-assembly of
particles with various morphologies \cite{He.2006, He.2008}.

In our previous publication we have shown that spinodal decomposition under
time-dependent quenches into the spinodal area reproduce experimental trends
\cite{meinArtikel1}. Therefore, solvent mixing in the present work is described
in a very similar manner. We model solvent mixing by a time dependent
interaction parameter between the solvent and the B-block, which is from now on
considered to be the solvent-phobic one. If not stated otherwise, its time
dependence is linear and given by 
\begin{equation}
\chi_{BS}(t)=
\begin{dcases*}
\chi_{BS}^{(0)} +s_{\chi}t, & $t\leq t_{max}$\\
\chi_{BS}^{(max)}, & $t>t_{max}$ \\
\end{dcases*}.
\label{eq: definition_lineare_zeitabhaengigkeit_chiBS}
\end{equation}
The cutoff time 
$t_{max}=\nicefrac{\left(\chi_{BS}^{(max)}-\chi_{BS}^{(0)}\right)}{s_{\chi}}$
can be interpreted as a mixing time, and the parameter
\begin{equation}
\chi_{BS}^{(0)}   = \frac{1}{2Nc_B f_P}+\frac{1}{2c_B(1-f_P)}
+\chi_{AB}c_A-\chi_{AS}\frac{c_A}{c_B}
\label{eq7: EPD_Spinodale}
\end{equation}
is the spinodal solvent-phobic interaction for which 
${\partial^2 F_{FH}}/{\partial f_P^2} = 0$, where $F_{FH}$ is the Flory-Huggins 
approximation to $F$ from equation (\ref{eq7: SCF_freie_Energie}) 
\cite{He.2006}. 

\section{SIMULATION SETUP}
\label{sec: simulation_setup}

Simulations start from randomly perturbed homogeneous initial states. All
simulation results are averaged over five simulation runs with different
initial conditions. We consider a volume fraction $f_P=0.1$ of a model polymer
with a solvent-philic block length $N_A=3$ and an incompatible solvent-phobic
block containing $N_B=14$ monomers. The incompatibility is described by an
interaction parameter $\chi_{AB}=1.05$, and to keep $\phi_A+\phi_B+\phi_S$
close to 1, the compressive modulus is set to $\kappa_H=100$. The mean volume fraction of
selective solvent is $f_S=1-f_P=0.9$. The diffusion coefficient $D_0$ in
equation (\ref{eq7: entwicklungsgleichung_omega}) can be substituted by $1$
without loss of generality as lengths are given in units of $l_0=R_g$ and times
in units of $t_0=\nicefrac{R_g^2}{D_0}$.  The solvent-philic interaction is
kept at a constant value $\chi_{AS}=-0.15$, and the solvent-phobic one is
varied from its spinodal value $\chi_{BS}^{(0)}=\chi_{BS}^{(Spin)}=1.249$ to
$\chi_{BS}^{(max)}=2.25$, which corresponds to a maximum quench depth of
approximately 1 like in reference \cite{meinArtikel1}. As in reference
\cite{meinArtikel1}, the random noise is turned off, i.e. $\eta_i = 0$.

The number of spatial grid points is set to $m=256\times 256$ with a lattice
constant $\Delta l=0.25 R_g$. The time step $h$ varies during the simulation as
described in \ref{sec: numerical_integration_scheme}. Unless stated otherwise,
we use an initial (and maximum) time step of $h=0.05 t_0$.  The segment length
in a polymer chain is $ds=\nicefrac{1}{N}$. Simulations are restricted to two
dimensions (2D) because the qualitative size dependence was shown to be
independent of the dimension in \cite{meinArtikel1}. In 2D, much larger systems
can be simulated over longer time periods.

The set of equations (\ref{eq7: diffusionsgl_g}) and (\ref{eq7:
bestimmung_phiA} -- \ref{eq7: bestimmung_phiS}) is solved with the same solvers
as in Ref.\ \cite{He.2006}. The evolution equations for the potential fields
(\ref{eq7: entwicklungsgleichung_omega}) is solved by a newly developed
semi-implicit integration scheme, which is presented in \ref{sec:
numerical_integration_scheme}.

\begin{figure*}[!ht]
\centering
\includegraphics[scale=1]{\dir/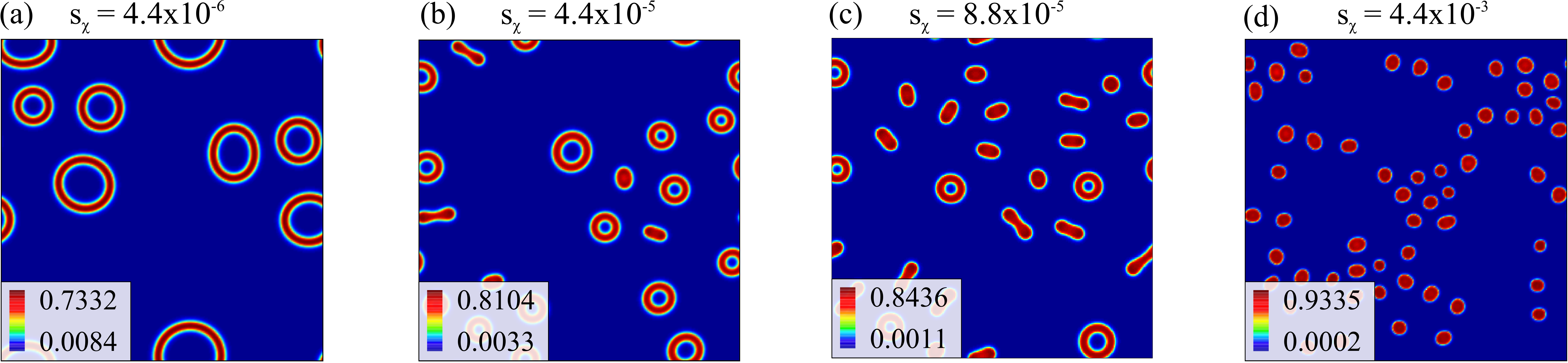} 
\caption{Color coded solvent-phobic density profiles $\phi_B$. (a) shows
simulation results at a mixing rate of $s_{\chi}=4.4\times 10^{-6}$ (b) at
$s_{\chi}=4.4\times 10^{-5}$, (c) at $s_{\chi}=8.8\times 10^{-5}$, and (d) at
$s_{\chi}=4.4\times 10^{-3}$. An increase of mixing rates does not only affect
particle sizes but also induces a morphological vesicle-to-micelle
transition.} \label{fig: morphologiebilder_simulations}
\end{figure*}

\section{RESULTS AND DISCUSSION}
\label{sec: results_and_discussion}

\subsection{Morphological transition and qualitative discussion of particle
size dependencies on mixing speeds}

Figure \ref{fig: morphologiebilder_simulations} depicts simulated polymer
particles for different mixing speeds $s_{\chi}$. Strictly speaking, only the
density of solvent-phobic B-monomers $\phi_B$ is shown, but since the
A-monomers accumulate approximately at green to yellow $\phi_B$-values, these
colors can be imagined to represent the solvent-philic A-block. It is evident
that an increase of the mixing rate leads to a decrease of the typical
particle size and induces a morphological vesicle-to-micelle transition: From
figure \ref{fig: morphologiebilder_simulations} (a) to (d), the number of
vesicles decreases while the number of micelles increases until only spherical
micelles are left.  For every $s_{\chi}$, the $\phi_B$-profiles in the early
stages of phase separation (not shown) closely resemble the profiles obtained
in our earlier work based on the Cahn-Hilliard equation for homopolymers
\cite{meinArtikel1}. 
Once the polymer content inside polymer aggregates is sufficiently high, the
block incompatibility leads to an internal rearrangement of copolymer chains
inside the aggregates, which eventually leads to the formation of different
morphologies. In the literature, different pathways of vesicle formation have
been described theoretically and observed experimentally
\cite{He.2006,Uneyama.2007,He.2008,weiss1, adams1,gummel1}. In the present
simulations, they form by nucleation and growth (mechanism II according to
Ref.\ \cite{He.2008}). The solvent-philic A-block, ultimately oriented towards
the solvent, sterically stabilizes the particles by suppressing further
ripening (where large particles grow and smaller particles dissolve) and
preventing ma\-cro\-phase separation \cite{He.2006}. 

\begin{figure}[!b]
\centering
\includegraphics[scale=1]{\dir/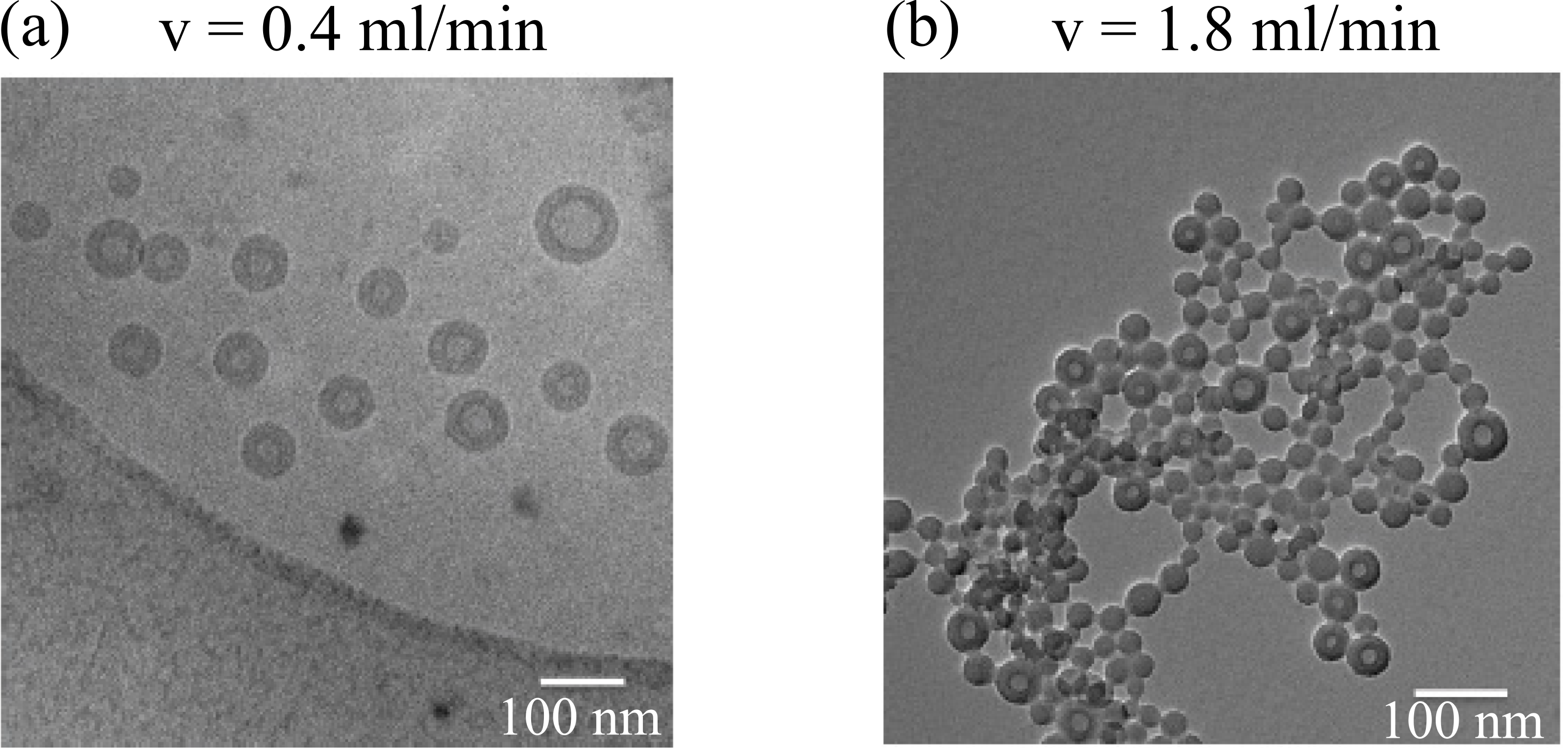} 
\caption{Transmission Electron Microscopy images of cross-linked nanoparticles
made of polybutadiene-polythyleneoxide diblock copolymers obtained in the
Caterpillar Micromixer at two different flow rates as indicated and otherwise 
identical experimental parameters under symmetric flow conditions.
After Ref.\ \cite{Thiermann.2014} (samples Cd10 and Cd11), 
courtesy of R. Thiermann.  }

\label{fig: morphologiebilder_experiments}
\end{figure}

An enrichment of micelles with increasing flow rate comparable to figure
\ref{fig: morphologiebilder_simulations} (a) to (c) is also observed
experimentally.  Transmission Electron Microscopy (TEM) images from experiments
are shown in figure \ref{fig: morphologiebilder_experiments}. Figure \ref{fig:
morphologiebilder_experiments} (a) and (b) provide a direct comparison of
identically prepared polymer solution samples for two different flow rates
\cite{Thiermann.2014} in the Caterpillar Micromixer \cite{Schonfeld.2004.2},
i.e. for two different mixing rates. The particle morphologies resemble the
simulation results from figure \ref{fig: morphologiebilder_simulations} (a) and
(c), except that the TEM images lack cylindrical micelles while the simulation
results contain a few. In other experimental work \cite{Mueller.2009}, such
cylindrical micelles have also been found to coexist with spherical micelles
and vesicles. Whether or not cylindrical micelles appear in simulations likely
depends on the interaction parameters $\chi_{AB}$ and $\chi_{AS}$. We conclude
that qualitatively, the dependence of morphologies on mixing rates is in good
agreement with experiments. Furthermore, the simulations indicate that the
enrichment of micelles only marks the onset of a complete morphological
transition from vesicular to micellar.

\subsection{Minkowski measures and determination of particle sizes}

\begin{figure*}[!t]
\centering
\includegraphics[scale=1]{\dir/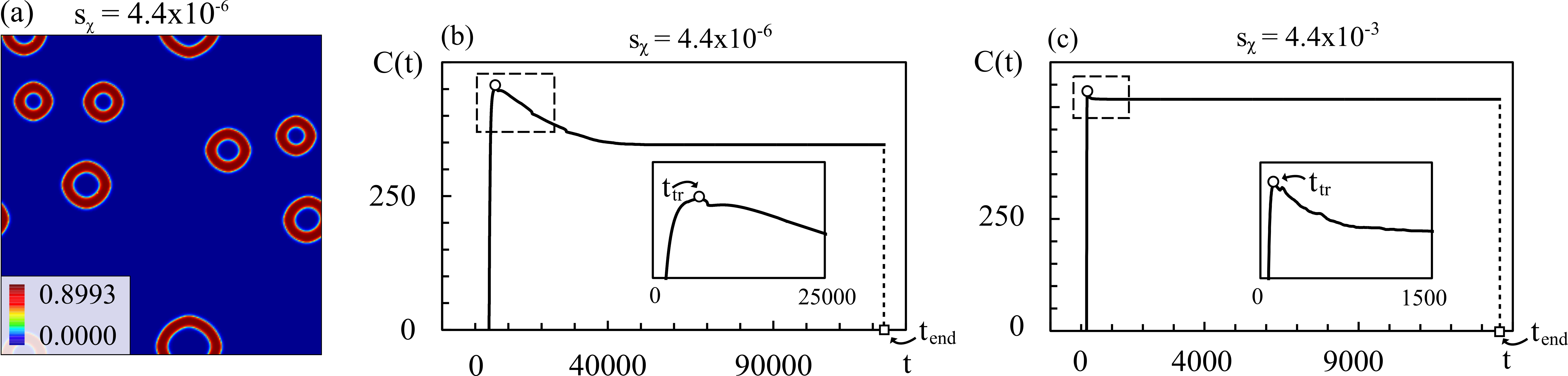} 
\caption{Vesicles at $t_{end}>t_{tr}$ (a) and time series of the Minkowski measure $C(t)$ for $s_{\chi}=4.4\times 10^{-6}$ (b) and $s_{\chi}=4.4\times 10^{-3}$ (c). The transition time $t_{tr}$ is labeled by a circle. $t_{end}$ denotes the end of a simulation run and is marked by a rectangle. The insets in (b) and (c) show a magnification of the respective part framed by the rectangle.}
\label{fig:minkowski_zeitreihen}
\end{figure*}

The snapshots in figure \ref{fig: morphologiebilder_simulations} are taken at
the so-called transition time $t_{tr}$ \cite{Sofonea.1999}. In the
Cahn-Hilliard model, the transition time separates a regime of spinodal
decomposition and initial droplet nucleation and growth, where polymer
aggregates form and concentrate, from a comparatively slow ripening regime,
where small aggregates grow at the expense of smaller ones until only one
single polymer aggregate remains. The transition time $t_{tr}$ can be
determined by inspecting, for instance, the Minkowski measure $C$
\cite{Sofonea.1999}. Here $C$ denotes the total boundary length of the union
over all subsets where $\phi_B$ exceeds a predefined threshold $\phi_B^{(th)}$.
The rapid concentration of solvent-phobic monomers in the spinodal
decomposition stage leads to a very fast temporal increase of $C$, while
ripening is characterized by a slow but steady decrease of $C$. This opposite
behavior leads to a clear maximum in $C(t)$, which marks the transition time.
Time series of $C$ in the SCF-EPD copolymer model are shown in figure
\ref{fig:minkowski_zeitreihen}.  They look very similar to the Cahn-Hilliard
model in our earlier studies \cite{meinArtikel1}. We still find the distinct
maximum that allows an analogous definition of the transition time $t_{tr}$. As
in the Cahn-Hilliard model, the steep rise at $t<t_{tr}$ is caused by a rapid
initial polymer aggregation, which is already associated with the formation of
different morphologies (see figure \ref{fig: morphologiebilder_simulations}).
The subsequent slow decrease of $C(t)$ is not caused by Ostwald Ripening,
\footnote{Although processes similar to Ostwald Ripening may contribute to a
small extent \cite{He.2006}.} but mainly by a shrinking of particles. In the
simulations from figure \ref{fig: morphologiebilder_simulations}, this
shrinking is most pronounced at $s_{\chi}=4.4\times 10^{-6}$. To give an
impression of the extent of the shrinking, figure
\ref{fig:minkowski_zeitreihen} (a) shows the density profiles from figure
\ref{fig: morphologiebilder_simulations} (a) at the end of the simulation run,
$t=t_{end}$. The shrinking and the increase of $\phi_B$-values between $t_{tr}$
and $t_{end}$ is most likely caused by the increase of $\chi_{BS}$ during this
time interval. Finally, $C(t)$reaches a plateau at late times (see figure
\ref{fig:minkowski_zeitreihen} (b) and (c)). This is in contrast to the
Cahn-Hilliard model, where $C(t)$ continues to decay at late times. The plateau
corresponds to a state where multiple stabilized particles are present. 

We measure particle sizes at transition time because the shrinking is not very
pronounced, and because a maximum of $C$ can be defined more precisely than
a plateau. To this end, pictures such as those shown in figure \ref{fig:
morphologiebilder_simulations} are converted into binary images with a
threshold value of $\phi_B^{(th)}$. Then a standard 4-connected-component image
labeling algorithm \cite{Nielsen.2014} is used to count and isolate single
particles. Afterwards, the area $A_i$ and circumference $U_i$ is determined for
every single particle $i=1,...,p$ in a picture with the Minkowski functional
algorithm from Mantz et al. \cite{Mantz.2008}. We define the sphere equivalent
radius $R_{s,i}$ and the vesicle equivalent radius $R_{v,i}$ of particle $i$ by 
\begin{equation}
R_{s,i}=\sqrt{\frac{A_i}{\pi}} \text{ and } R_{v,i}=\frac{A_i}{U_i}+\frac{U_i}{4\pi},
\end{equation}
respectively. The vesicle equivalent radius is the outer radius of the
spherical shell with area $A_i$ and total perimeter $U_i$. In case a spherical
micelle is processed, $R_{s,i}$ and $R_{v,i}$ are equal to its radius $R$,
which can be verified by insertion of $A_i=\pi R^2$ and $U_i=2\pi R$. Mean
particle sizes are estimated by the mean values
\begin{equation}
R_s=\frac{1}{p}\sum_{i=1}^{p} R_{s,i} \text{ and } R_v=\frac{1}{p}\sum_{i=1}^{p} R_{v,i}.
\label{eq: partikelgroessen}
\end{equation}
As a measure for polydispersity of a nanoparticle population we use the
standard deviation 
\begin{equation}
\Delta R_{j} = \sqrt{\frac{1}{p}\sum_{i=1}^{p}\left(R_j - R_{j,i}\right)^2}. 
\label{eq: partikelgroessenvarianz}
\end{equation}
for $j=s,v$.

\subsection{Simulation results for particle sizes and mor\-pho\-lo\-gi\-cal
re\-gi\-mes}

\begin{figure}[!b]
\centering
\includegraphics[scale=0.9]{\dir/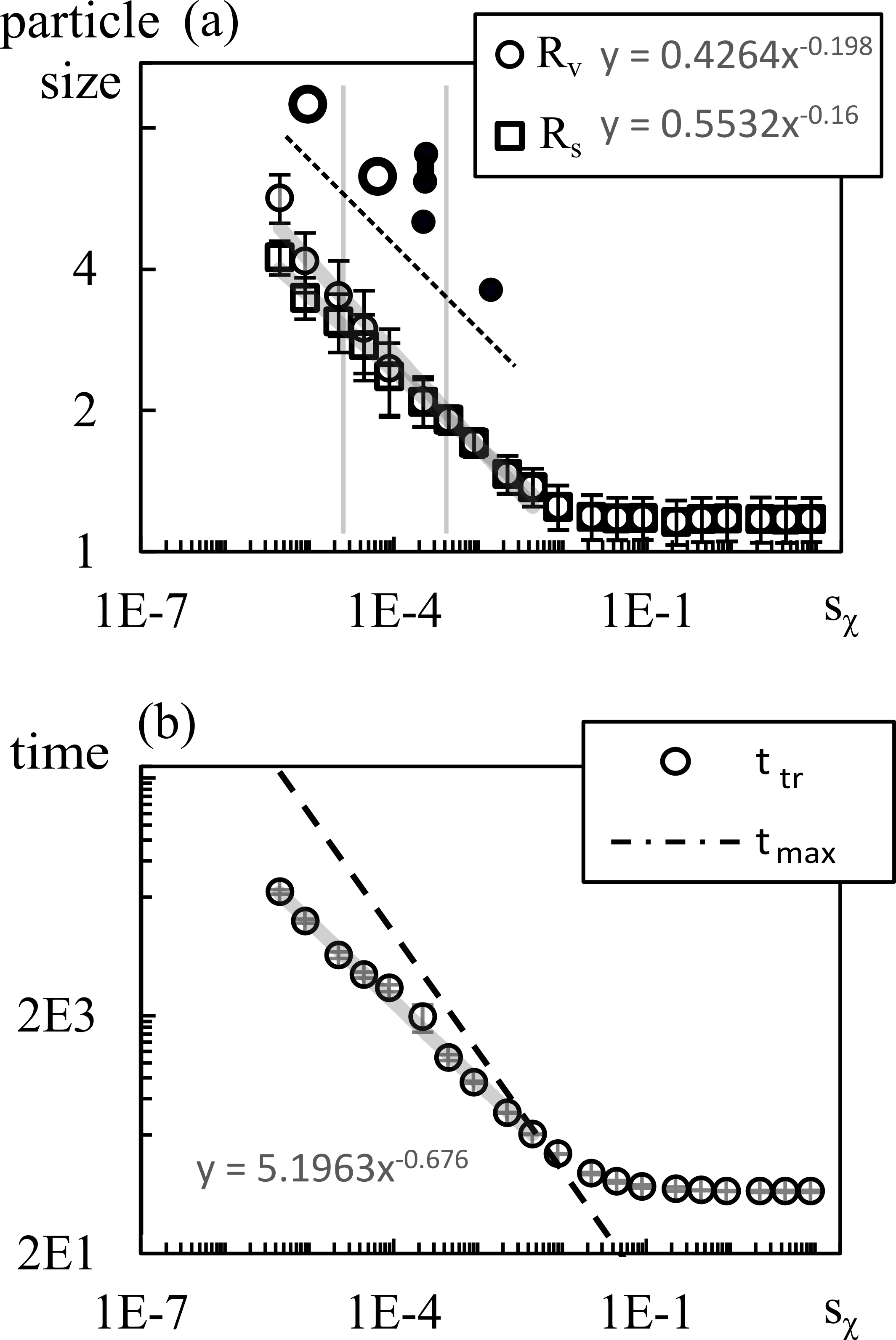} 
\caption{Simulation results for particle sizes (a) and transition times (b) in
dependence on the quench rate $s_{\chi}$. The data points in (a) represent the
simulation results for the sphere and vesicle equivalent radius from equation
(\ref{eq: partikelgroessen}) and the error bars are the standard deviation from
equation (\ref{eq: partikelgroessenvarianz}). Trend line equations are given in
the legend. The vertical lines separate three different morphological regimes,
where the corresponding morphologies above the dashed line appear. Empty
circles symbolize vesicles, solid black dots spherical micelles and the
dumbbell cylindrical micelles. The data points in (b) are the transition times
$t_{tr}$ (cp. figure \ref{fig:minkowski_zeitreihen}) and the dashed line is
$t_{max}$ from equation (\ref{eq: definition_lineare_zeitabhaengigkeit_chiBS}).
The trend line equation refers to the transition time. Error bars are standard
deviations over five different simulation runs. } 
\label{fig: rate-size-relations_linear}
\end{figure}

Figure \ref{fig: rate-size-relations_linear} shows simulation results for
particle sizes and transition times\footnote{In the present article we only use
geometric quantities to determine particle sizes. Due to the formation of
particle clusters like in figure \ref{fig: morphologiebilder_simulations} (b)
and (d), for instance, structure factors are multimodal. Therefore, extracting
particle sizes from the structure factor (e.g., from the first moment or the
maximum) is difficult.}. Particle sizes are given in units of the polymer's
radius of gyration $l_0 = R_g$ and times in units of
$t_0=\nicefrac{R_g^2}{D_0}$. The dependence of particle sizes and transition
times on quench rates also resembles the results from the Cahn-Hilliard model
for homopolymers: There is an asymptotic regime, where particle sizes converge
to results for a constant quench depth $\chi_{BS}(t)\equiv\chi_{BS}^{(max)}$,
and there is a non-asymptotic regime, where particle sizes follow scaling laws
$R \propto s_{\chi}^{\alpha}$ with $\alpha\approx -\nicefrac{1}{6}$, while
transition times can be approximated by $t_{tr}\propto
s_{\chi}^{-\nicefrac{2}{3}}$ \cite{meinArtikel1}. As in reference
\cite{meinArtikel1}, the asymptotic and the non-asymptotic regime are separated
by the time when $t_{max}$ in figure \ref{fig: rate-size-relations_linear}
intersects $t_{tr}$.  The similarity of the data curves in figure \ref{fig:
rate-size-relations_linear} to our earlier Cahn-Hilliard results for
homopolymers \cite{meinArtikel1} confirms the hypothesis that particle sizes
are determined during the very early stages of phase separation, where the
specific sequence of polymers (block copolymers vs.\ homopolymers) is not yet
important. In particular, the predictions of the simple Cahn-Hilliard model
\cite{meinArtikel1} can still be applied. 


For diblock-copolymers, however, the non-asymptotic re\-gime splits into three
different morphological regimes. We call a morphological regime an interval of
mixing rates with specific predominant particle morphologies. In figure
\ref{fig: rate-size-relations_linear} (a) the morphological regimes are
separated by the vertical lines. At low mixing rates only vesicles form.
Figure \ref{fig: morphologiebilder_simulations} (a) shows a corresponding
snapshot of particles. At intermediate mixing rates one gets a mixture of
vesicles, cylindrical micelles and spherical micelles. In the intermediate
regime the number of vesicles decreases with increasing mixing rates while the
number of micelles increases. Corresponding snapshots of particles can be found
in figure \ref{fig: morphologiebilder_simulations} (b) and (c). At large mixing
rates there are only micelles as seen in figure 
\ref{fig: morphologiebilder_simulations} (d). Because every morphological 
regime covers a certain interval of particle sizes, particle morphologies are
directly coupled to their size. With respect to the co-solvent method this
means that it should be possible to produce pure vesicular or micellar
populations at 'extremely' large or low flow rates, but intermediate ones
typically result in heterogeneous morphologies. 
As long as the micelles are not exclusively spherical, $R_v$ is larger than $R_s$ because the
formation of vesicles from a polymer aggregate increases the perimeter $U_i$
of a spherical shell, while it keeps the volume $A_i$ constant.

It should be noted though that the simulated ''micelles'' from the rightmost
morphological regime in figure \ref{fig: rate-size-relations_linear} (a) are,
strictly speaking, actually no 'real' micelles. Real micelles are equilibrium
structures with a well-defined size distribution and composition that does
not depend on the history of the system. Here, we consider 'micelle-like'
spheres with the typical core-shell structure that would be characteristic
for a micelle, but they are not equilibrated.

%

\subsection{Comparison of particle sizes to experimental results}

\begin{figure}[!b]
\centering
\includegraphics[scale=0.9]{\dir/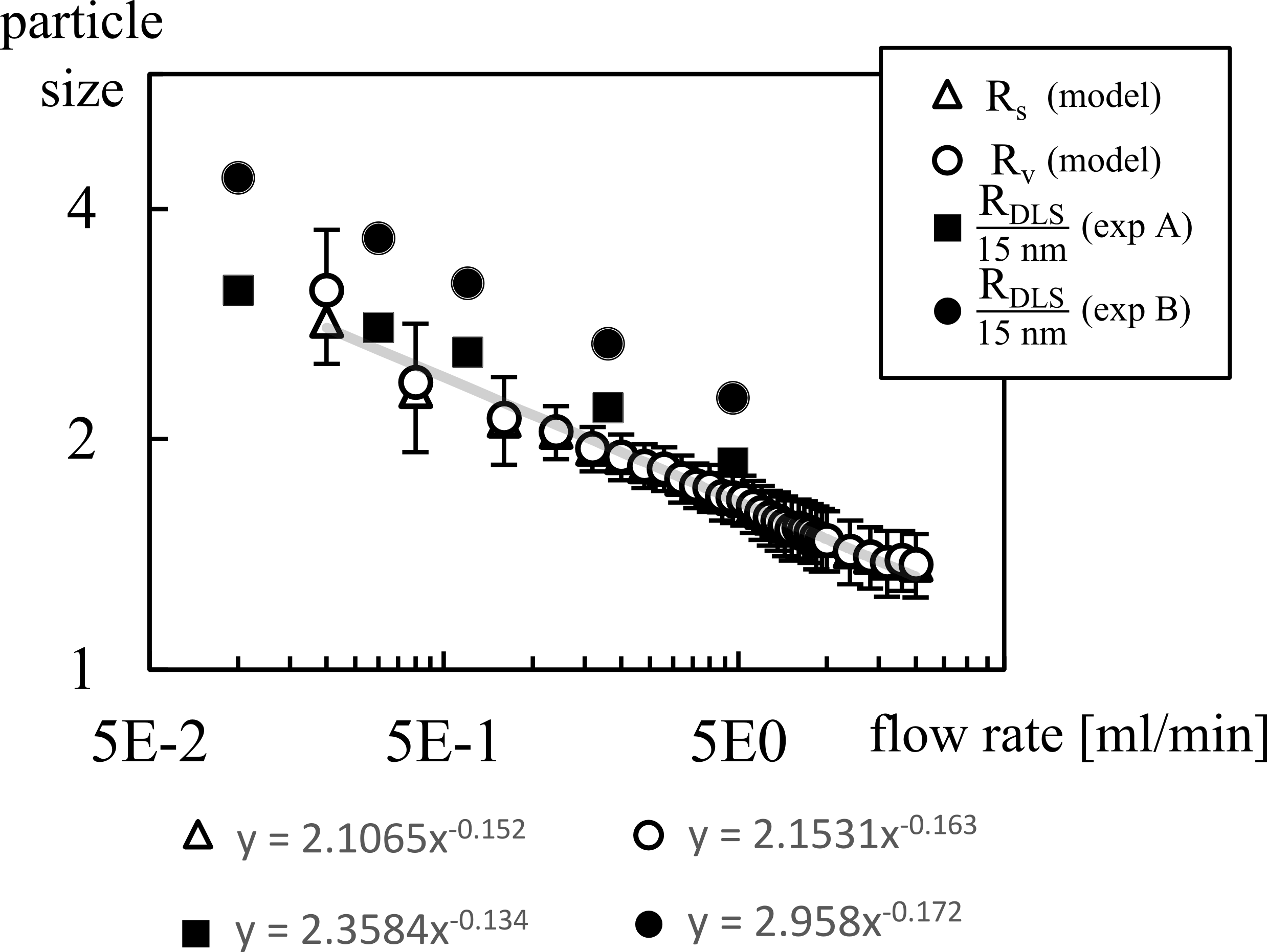} 
\caption{Particle sizes in units of $R_g$ versus flow rates $v$ in ml/min in
the Caterpillar Micromixer. The open symbols represent simulation results and
the solid symbols experimental results from Thiermann et al.
\cite{Thiermann.2012}. A ($\hat{=}$ H) and B
($\hat{=}\text{CO}\text{--CH}_2\text{--CH}_2\text{--COOH}$ ) denote different
end groups attached to the polymer. Trend line equations are shown beneath the
diagram at the corresponding symbols. For the sake of simplicity only the
regression line to $R_s$ is shown (thick light grey line). Error bars mark 
the polydispersity as determined from Eq.\ \ref{eq: partikelgroessenvarianz} 
and are again only shown for $R_s$ for the sake of clarity. $\Delta R_v$ 
looks very similar.}
\label{fig: CPMM_partikelgroessenvergleich}
\end{figure}


After having studied the idealized situation where the solvent mixing is
described by the simple linear increase of $\chi_{BS}(t)$ with time, (Eq.\
(\ref{eq: definition_lineare_zeitabhaengigkeit_chiBS})), we will now
introduce a more general Ansatz which allows us to compare simulation
results directly to experiments: We assume that $\chi_{BS}(t)$ is a linear
function of the volume fraction of selective solvent $\varphi_{SS}(t)$, which
we take to be homogeneous throughout the system. Specifically, we assume
\begin{equation}
\chi_{BS}(t)=\chi_{BGS}+\frac{\chi_{BSS}-\chi_{BGS}}{1-f_P}\varphi_{SS}(t),
\label{eq: chi_in_mikromischern}
\end{equation} 
where $\chi_{BGS}$ is the interaction parameter between the B-monomers and the
good solvent, while $\chi_{BSS}$ describes the interaction of the B-monomers
and the selective solvent. The Ansatz ensures that $\chi_{BS}=\chi_{BGS}$ if
no selective solvent is present and $\chi_{BS}=\chi_{BSS}$ if only selective
solvent is present, i.e. if $\varphi_{SS}=1-f_P $. 

We assume that $\varphi_{SS}(t)$ can be calculated independently of the
particle self-assembly simply by considering the mixing of simple fluids in a
given micromixer geometry.  This Ansatz allows a direct coupling of mixer
geometries and flow rates into the description of particle growth. Here we
specifically consider the case of the Caterpillar Micromixer (CPMM), where
analytical expressions for $\varphi_{SS}(t)$ have been derived by Schoenfeld et
al.\ \cite{Schonfeld.2004.2}.  Simulation results for particle sizes with the
mixing profile $\varphi_{SS}(t)$ from the Caterpillar micromixer are shown in
figure \ref{fig: CPMM_partikelgroessenvergleich}. Even with this Ansatz, the 
data still show scaling behavior. 

If we normalize the experimental data with 15 nm, i.e., we assume the gyration
radius of the polymers to be around $R_g=$15 nm, we get almost quantitative
agreement between simulation data and experiments. In reality, the radius of
gyration is probably smaller, in the range of 5-10 nm \cite{meinArtikel1,
simon_diss}, hence the simulations probably underestimate the particle size. 
The simulations also reproduce the experimentally observed morphological
transition from vesicles to micelles upon reducing the flow rate, but here
again, the absolute values of flow rates where the morphological transition is
observed in experiments and simulations do not match quantitatively.  In the
experiments pure vesicular populations are observed up to flow rates of
approximately 1.8 ml/min, while in the simulations, micelles dominate already
at flow rates above approximately 0.8 ml/min. The specific flow rate values of
the vesicle to micelle transition as well as the final particle sizes are
likely to depend strongly on $\chi_{AB}$ and $\chi_{AS}$.  Since the material
parameters of the experimental systems studied by Thiermann et al.\
\cite{Thiermann.2012, Thiermann.2014} were not determined, a direct
quantitative comparison between model and experiments is difficult.

Nevertheless, our results show that the simulations reproduce the experimental
trends and the order of magnitude of simulated particle sizes matches the
experimentally observed sizes. This suggests that the main mechanism
determining particle sizes is captured by our mean field model, even though it
does not account for Brownian motion of particles and thereby induced 
collision induced coagulation and growth of particles. 

To conclude, the results show that the combination of established
methods for calculating solvent mixing in complex geometries on large scales
with mean field theories for aggregation on mesoscales, coupled through an
Ansatz like equation (\ref{eq: chi_in_mikromischern}), represent a promising
multiscale approach for describing flow controlled particle assembly in
micromixers.

\section{SUMMARY AND OUTLOOK}
\label{sec: summary_and_outlook}

We have implemented time dependent interaction parameters into a Dynamic Self
Consistent Field Theory for copolymer simulations in order to describe the
nonequilibrium assembly of amphiphilic diblock-copolymers in micromixers.

Experimental observations show that the final morphologies of particles are
mostly vesicular at low mixing rates, and mostly micellar at high mixing rates
\cite{Mueller.2009, Thiermann.2012, Thiermann.2014}. Our simulations indicate
that these changes in morphologies are the signature of an incomplete
vesicle-to-micelle transition, and that nanoparticle populations should
be purely micellar if the flow rates are sufficiently large. We conclude
that it should be possible to produce both pure vesicular and pure micellar 
nanoparticle populations with the co-solvent method. In the morphological
regime at low flow rates, one can tune the size of vesicles, while in the regime
at high flow rates the size of micelles can be adapted. The intermediate
regime, however, excludes certain mean sizes if one wants to produce particles
with only one morphology. 

The rate of solvent mixing qualitatively affects particle sizes in the same way
in the SCF-EPD model as in the Cahn Hilliard model for homopolymers. The
present work hence confirms our conjecture in Ref.\ \cite{meinArtikel1}, that
the sizes of particles that are aggregated from amphiphilic copolymer solutions
are determined during the very early stages of phase separation by a
competition between the interfacial tension of diffuse interfaces and the
decreasing solvent quality for the solvent-phobic block. Furthermore, figure
\ref{fig: rate-size-relations_linear} shows that the rate-dependent
morphological changes do not affect the typical scaling behavior $R \appropto
v^{-\nicefrac{1}{6}}$ in the non-asymptotic regime, which is observed here in
copolymer solutions as well as in Ref.\ \cite{meinArtikel1} in homopolymer
solutions. 

As we have mentioned above, vesicles form via a nucleation-and-growth pathway
in the simulations considered here (pathway II in Ref.\ \cite{He.2006}), which
is typical for the aggregation of particles from solutions with relatively low
copolymer densities and/or relatively weak incompatibilities \cite{He.2008}. At
higher copolymer concentrations \cite{He.2008,gummel1}, one observes an
alternative aggregation-and-bending pathway (pathway I), where micellar
particles first aggregate to platelets and these platelets then bend around,
driven by a competition of bending energy and line tension, to form closed
vesicles. In future work, it will be interesting to study the influence of the
mixing rate on particle aggregation in the regime where pathway I is dominant.
(These simulations would have to be done in three dimensions, since in two
dimensions, the final step of platelet bending does not take place). In
particular, it will be interesting to investigate whether the scaling law for
the particle sizes, $R\appropto s_{\chi}^{-\nicefrac{1}{6}}$, still persists in
that regime. If the scaling law is destroyed, this would provide indirect
evidence that the vesicle formation in the micromixers that have studied
experimentally proceeds predominantly via mechanism II.

To implement mixer geometries into solvent mixing, we have coupled an
established description for the Caterpillar Micromixer (CPMM)
\cite{Schonfeld.2004.2} into the SCF-EPD model. Here, the interaction between
the solvent and the solvent-phobic monomers was assumed to be a linear function
of the volume fraction of selective solvent (cf.\ Eq.\ \ref{eq:
chi_in_mikromischern}). This Ansatz allowed us to perform simulations of
particle assembly with realistic mixing times and to compare directly the
particle sizes obtained in experiments and simulations. The scale of the
simulated particle sizes matches experimentally determined particle sizes, and
the scaling behavior $R\appropto v^{-\nicefrac{1}{6}}$ was also reproduced,
which implies that the mean field theories are suited to capture the 
self-assembly process in the co-solvent method.

More generally, our Ansatz represents a computationally very efficient
multiscale approach to describe the nanoparticle formation in mixer geometries
on millimeter or centimeter scales over realistic mixing times. It is not
restricted to the CPMM geometry, other micromixers can be implemented as well -
even though, in most cases, the resulting mixing behavior would have to be
calculated numerically if approximate analytical solutions are not available.
The mesoscale model can also be refined. For example, it can can be extended
such that it allows for spatial variations of the selective solvent volume
fraction $\phi_{SS}(t)$ \cite{simon_diss}. In the present work, we have assumed
$\phi_{SS}=\varphi_{SS}$ to be constant everywhere; in reality, one would
expect that the selective solvent preferably accumulates outside of the
particles, and this may influence the particle assembly. Extending the model to
more complex copolymer architectures, and to copolymer mixtures, is also
possible and quite straightforward. In the future, it will be particularly
challenging to use our methods to model the self-assembly of complex structured
nanoparticles.



We should however, also point out some limitations of the approach. The main
advantage of using mean field continuum theories like the SCF-EPD model rather
than detailed particle models is the possibility to simulate self-assembly on
realistic mixing times in micromixers. This comes at the prize that
the simulations lack detail on a molecular level. The SCF-EPD model makes a
local equilibrium assumption and thus cannot capture situations where chains
freeze or crystallize partially and chain conformations can no longer
equilibrate.  The scaling law $R\propto s_{\chi}^{-\nicefrac{1}{6}}$ is typical
for liquid-liquid demixing as particle sizes are reported to depend differently
on mixing rates when particle growth interferes with solidification
\cite{Franeker.2015}.  Hence the SCF-EPD model should only be applied if
particles initially form during liquid-liquid demixing.  

Furthermore, so far, we have not investigated the direct effect of flow on
self-assembly.  Although the size control appears to be dominated by solvent
mixing rates, flow effects might still affect the particle size dependence on
flow rates to some degree. This is particularly relevant in shear flows at high
shear rates.  In a recent study it was found that shear flow in micro channels
may delay nucleation \cite{JohannesArtikel} and thus, growing shear rates could
directly increase the transition times $t_{tr}$. Furthermore, it was found that
strong shear flows may induce irreversible changes in the final shapes of
particles.  Studying the effect of solvent mixing in the presence of shear flow
by coupling solvent mixing into the hybrid EPD-LB model by Heuser et al.
\cite{JohannesArtikel} will be an interesting topic for future work as well.
If the shear rate is so high that it becomes comparable to the inverse Zimm
time of polymers, polymers will deform, which will further influence the
self-assembly. However, in micromixers, such high shear rates are usually not
applied.

%

\section*{ACKNOWLEDGMENTS}

We thank R. Thiermann for allowing us to reproduce the experimental images in
figure \ref{fig: morphologiebilder_experiments} from his thesis,
\cite{Thiermann.2014}. We acknowledge funding by EFRE (Europ\"aischer Fonds
f\"ur regionale Entwicklung) and by the German Science Foundation within the
Collaborative Research Center TRR 146. The simulations were carried out on the
high performance computing center MOGON at JGU Mainz. FS wishes to thank Axel
H.E. M\"uller and Andr\'e Gr\"oschel for their inspiring work and enjoyable
discussions on complex nanoparticle assembly.

\appendix

\section{NUMERICAL INTEGRATION SCHEME}
\label{sec: numerical_integration_scheme}

\begin{figure*}[!t]
\centering
\includegraphics[scale=1]{\dir/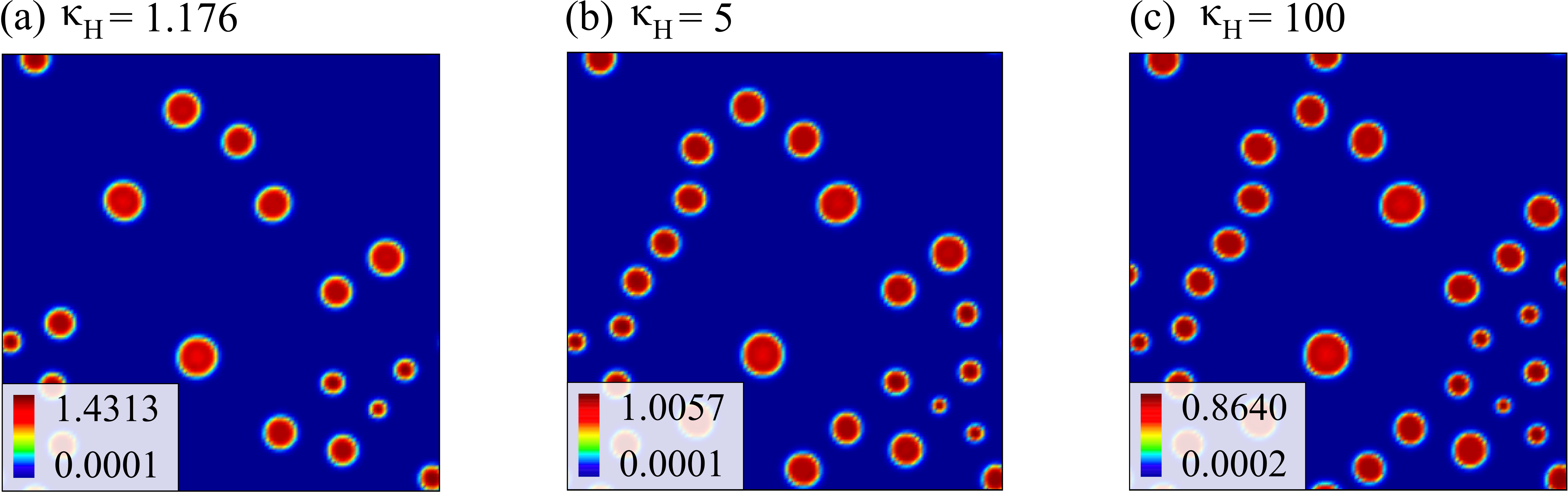} 
\caption{Color coded two-dimensional solvent-phobic density profiles $\phi_B$
of stable micelles for different compressive moduli $\kappa_H=1.176$ (a),
$\kappa_H=5$ (b) and $\kappa_H=100$ (c). Simulations were performed on a
$128\times 128$ grid with a lattice constant $\Delta l=\nicefrac{1}{3}$ and a
fixed time step of $h=0.002$. The values of the remaining parameters are
$\chi_{BS}=1.6$, $f_P=0.1$, $N_A=2$, $N_B=15$, $\chi_{AB}=1.05$,
$\chi_{AS}=0.0375$, $D_P=\nicefrac{1}{N}$, and $D_S=1$. All snapshots are taken
at time step $n=200000$.}
\label{fig: Kompressibilitaetsvergleich}
\end{figure*}

The value of the compressive modulus $\kappa_H$ affects the number, the mean
size, and the polymer content of simulated nanoparticles as shown in figure
\ref{fig: Kompressibilitaetsvergleich}. 

\begin{figure*}[!ht]
\begin{minipage}[!t]{1.0\textwidth} 
\hrule
\vspace{2mm}
$$
\left[ 1-hD_P\left(  (1+2\tilde{\kappa}_H\phi_A^{(n)})\Delta - 4\beta_A\tilde{\kappa}_H\phi_A^{(n)}\nabla\omega_A^{(n)}\cdot \nabla \right) \right] \omega_A^{(n+1)} 
- \left[hD_P \tilde{\kappa}_{H} 2\phi_B^{(n)}\left(\Delta-2\beta_B\nabla\omega_B^{(n)}\cdot \nabla\right)\right]\omega_B^{(n+1)}
$$
\begin{equation}
-\left[hD_P \tilde{\kappa}_{H}\phi_S^{(n)} (\Delta -\beta_S\nabla\omega_S^{(n)}\cdot \nabla)\right]\omega_S^{(n+1)} = \omega_A^{(n)}-hD_P E_A^{(n)},
\label{eq7: neues_schema_omega_A}
\end{equation}

$$
-\left[hD_P \tilde{\kappa}_{H}2\phi_A^{(n)}\left(\Delta-2\beta_A\nabla\omega_A^{(n)}\cdot\nabla\right)\right]\omega_A^{(n+1)}
+\left[ 1 - hD_P \left(  (1+2\tilde{\kappa}_H\phi_B^{(n)})\Delta -4\beta_B\tilde{\kappa}_H\phi_B^{(n)}\nabla\omega_B^{(n)}\cdot \nabla  \right)\right]\omega_B^{(n+1)}
$$
\begin{equation}
-\left[ hD_P \tilde{\kappa}_{H}\phi_S^{(n)} \left(\Delta-\beta_S\nabla\omega_S^{(n)}\cdot \nabla\right) \right]\omega_S^{(n+1)} = \omega_B^{(n)} - hD_P E_B^{(n)},
\label{eq7: neues_schema_omega_B}
\end{equation}

$$
- \left[ hD_S \tilde{\kappa}_{H} 2\phi_A^{(n)}\left(\Delta -2\beta_A\nabla\omega_A^{(n)}\cdot\nabla\right)\right]\omega_A^{(n+1)}
- \left[ hD_S \tilde{\kappa}_{H}2\phi_B^{(n)}\left(\Delta -2\beta_B\nabla\omega_B^{(n)}\cdot \nabla\right) \right]\omega_B^{(n+1)}
$$
\begin{equation}
+\left[1-hD_S \left( (1+\tilde{\kappa}_H\phi_S^{(n)})\Delta - \beta_S\tilde{\kappa}_H\phi_S^{(n)} \nabla\omega_S^{(n)}\cdot \nabla\right)\right]\omega_S^{(n+1)} = \omega_S^{(n)} - hD_S E_S^{(n)}.
\label{eq7: neues_schema_omega_S}
\end{equation}
\hrule
\end{minipage} 
\caption{Semi-implicit integrator for the potential field equations \ref{eq7:
entwicklungsgleichung_omega} to increase stability regions at high $\kappa_H$.
\label{fig: semiimpliziter_Integrator}
}
\end{figure*}

Polymer solutions typically possess a liquid-like compressive modulus
\cite{Freed.1995}, so simulations should be per\-formed at high $\kappa_H$. 
If the chemical potentials $\mu_i$ from equations \ref{eq7:
variationsableitung_F_phi_A_EPD} -- \ref{eq7: variationsableitung_F_phi_S_EPD}
are inserted into the dynamical equations \ref{eq7:
entwicklungsgleichung_omega}, $\kappa_H$ appears as a prefactor of Laplacian
terms $\Delta\phi_i$. Hence, the stiffness of the dynamical equations increases
with $\kappa_H$.  Stiffness of differential equations can be tackled by 
applying implicit or semi-implicit numerical integration sche\-mes
\cite{Dahlquist.1963, Zhu.1999, Ascher.1995, Uneyama.2007}.

A standard approach \cite{Dahlquist.1963,Zhu.1999,Ascher.1995} to cope with the
stiffness caused by high $\kappa_H$ in equation (\ref{eq7:
entwicklungsgleichung_omega}) is to carry out a direct implicit quadrature of
the term $\int_{t_n}^{t_{n+1}}\kappa_H\left(\Delta \phi_A(t)+\Delta
\phi_B(t)+\Delta \phi_S(t)\right) \,dt$ up\-on integrating the equation over a
time step $h=t_{n+1}-t_n$. Hence, the right hand side of the time discrete
version of equation (\ref{eq7: entwicklungsgleichung_omega}) depends on
$\phi_i^{(n+1)}$, which in turn depends on $\omega_i^{(n+1)}$ via the
(discretized) equations (\ref{eq7: diffusionsgl_g}) and  (\ref{eq7:
bestimmung_phiA} -- \ref{eq7: bestimmung_phiS}). Here $\phi_i^{(n)} \approx
\phi_i(t_n)$ and $\omega_i^{(n+1)} \approx \omega_i(t_{n+1})$ describe the
discrete evolution of $\phi_i, \omega_i$ in the time discrete dynamics.  In
order to solve equation (\ref{eq7: entwicklungsgleichung_omega}) for
$\omega_i^{(n+1)}$, one must thus solve the whole complexly nested non-linear
system of the discretized equations (\ref{eq7: diffusionsgl_g}), (\ref{eq7:
bestimmung_phiA} -- \ref{eq7: bestimmung_phiS}), and (\ref{eq7:
entwicklungsgleichung_omega}) by an interative method.  The unknowns of the
SCF-EPD model are $\omega_i^{(n+1)}(\vec{r})$, $\phi_i^{(n+1)}(\vec{r})$,
$g^{(n+1)}(\vec{r},s)$, and $g'^{(n+1)} (\vec{r}, s)$ with $i=A,B,S$ at all $m$
spatial grid points and discretized positions $s=n\,ds$ of distance
$ds=\nicefrac{1}{N}$ along a polymer chain with $n\in\{0,...,N\}$. It is
evident that the number of unknowns and thus, the dimension of the discretized
system of equations, increases dramatically with the polymer chain length $N$.
In fact, the iteration along the polymer chain for every $\vec{r}$ to calculate
$g$ and $g'$ typically consumes by far the most part of the computation time
even for \textit{algebraic} update rules from explicit integrators for equation
(\ref{eq7: diffusionsgl_g}). Therefore, solving the complete set of equations
(\ref{eq7: diffusionsgl_g}), \ref{eq7: bestimmung_phiA} -- \ref{eq7:
bestimmung_phiS}, and \ref{eq7: entwicklungsgleichung_omega} by an
\textit{iterative} method would lead to a particularly dramatic increase of
simulation times with growing $N$. 

Hence, implicit iterative schemes that involve multiple calculations of
$\phi_{i}^{(n+1)}$ help to overcome stiffness instabilities at high values of
$\kappa_H$, but they come at the cost of a dramatic increase in computation
times. To avoid this problem while still enlarging the $\kappa_H$-region where
the the algorithm runs stably, we have developed a semi-implicit integrator for
equation (\ref{eq7: entwicklungsgleichung_omega}) that does not require the
computation of $\phi_i^{(n+1)}$. The main equations are summarized in figure
\ref{fig: semiimpliziter_Integrator} (equations (\ref{eq7:
neues_schema_omega_A}--\ref{eq7: neues_schema_omega_S})), the derivation will
be shown below. In this scheme, the potential fields
$\omega_i^{(n+1)}(\vec{r})$ are decoupled from the other unkown variables
$\phi_i^{(n+1)}(\vec{r})$, $g^{(n+1)}(\vec{r},s)$, and $g'^{(n+1)}(\vec{r},s)$.
This decoupling allows one to selectively deal with $\kappa_H$-induced
stiffness of the dynamical equations while keeping the computational effort per
time step as close to efficient explicit schemes as possible. In other words,
it guarantees that, in particular, $g^{(n+1)}(\vec{r},s)$ and
$g'^{(n+1)}(\vec{r},s)$ can still be calculated by cost-efficient explicit
sche\-mes, whereas the iterative methods are only applied to a
$3m$-di\-men\-sio\-nal system of equations for $\omega_i(\vec{r},t_{n+1})$. At
$\kappa_H=100$, our semi-implicit integration scheme allows us to use much
larger time steps than, e.g., the explicit scheme used in earlier work
\cite{He.2006}, and as a result, the simulations are up to 100 times faster
\cite{simon_diss}.  We solve equation (\ref{eq7: diffusionsgl_g}) with the
explicit scheme from Tzemeres et al.  \cite{Tzeremes.2002} and to calculate
$\phi_i$, the integrals in equations (\ref{eq7: bestimmung_phiA} -- \ref{eq7:
bestimmung_phiS}) are approximated by a standard Euler method. 

The first step in the derivation of this integrator from equation (\ref{eq7:
entwicklungsgleichung_omega}) is to extract explicit $\omega_i$-ex\-press\-ions
from $\Delta \phi_i$.  Equation (\ref{eq7: bestimmung_phiS}) directly yields
the exact relation
\begin{equation}
\Delta\phi_S = \phi_S \left[ \nabla\omega_S \cdot\nabla\omega_S - \Delta \omega_S\right].
\end{equation}
To obtain analogous expressions for $\Delta\phi_A$ and $\Delta\phi_B$ we apply
the Feynman-Kac formula. It states that solutions to equation (\ref{eq7:
diffusionsgl_g}) can be defined recursively by
\begin{equation}
\begin{aligned}
\begin{split}
g[\omega](\vec{r},s)=& \exp\left(-ds N\omega(\vec{r})\right) \int_V \Phi(\vec{r}-\vec{r}\,')g[\omega](\vec{r}\,',s-ds) d\vec{r}\,'
\\  \eqqcolon & \exp\left(-ds N\omega(\vec{r})\right)\, I[\omega](\vec{r}) ,
\end{split}
\end{aligned}
\label{eq: Feynman-Kac_g}
\end{equation}
where $\Phi$ is proportional to the bond transition probability of a Gaussian
chain \cite{Fredrickson.2006} and $\omega$ is given by equation (\ref{eq7:
omega_g}). Likewise, one has
\begin{equation}
g'[\omega](\vec{r})=\exp(-ds N \omega)\, I'[\omega](\vec{r})
\label{eq: Feynman-Kac_g'}
\end{equation}
with $\omega$ from equation (\ref{eq7: omega_g'}). Inserting the recursive
definitions of the end segment distribution functions from equations (\ref{eq:
Feynman-Kac_g}) and (\ref{eq: Feynman-Kac_g'}) into the respective equation
(\ref{eq7: bestimmung_phiA}) or (\ref{eq7: bestimmung_phiB}) yields
\begin{equation}
\Delta \phi_i  = \phi_i\left[4 ds N \nabla\omega_i\cdot\nabla\omega_i -2dsN\Delta\omega_i\right] + R_i 
\end{equation}
for $i=A\text{, }B$. The remainder $R_i$ summarizes all non-leading stiffness
contributions, which in this case are all terms that contain derivatives of the
integrals $I$ and $I'$ with respect to the entries $r_j$ of
$\vec{r}=(r_1,r_2,r_3)^T$. Since we use $ds=\nicefrac{1}{N}$ in our simulations,
we set $dsN=1$. In the following we use the short hand notations 
\begin{equation}
X_S=\Delta\omega_S-\beta_S\nabla\omega_S\cdot\nabla\omega_S
\label{eq: X_S_abkuerzung}
\end{equation}
and
\begin{equation}
X_i=2\Delta\omega_i-\beta_i4\nabla\omega_i\cdot\nabla\omega_i
\label{eq: X_AB_abkuerzung}
\end{equation}
for $i=A\text{, }B$, where $\beta_A$, $\beta_B$, $\beta_S\in[0,1]$ are damping
coefficients to adjust the 'degree' of the implicit treatment and to regulate
truncation errors: the smaller the damping coefficients, the smaller are
truncation errors. To prepare the dynamical equations (\ref{eq7:
entwicklungsgleichung_omega}) for the semi-implicit integrator we define
$\tilde{\kappa}_H=\alpha\kappa_H$ with another damping coefficient
$\alpha\in[0,1]$ and zero-pad their time integrals according to 
\begin{equation}
\begin{aligned}
\begin{split}
\omega_i(t_{n+1}) & =\, \omega_i(t_n) - D_i \int_{t_n}^{t_{n+1}}\Big\{
\Delta\mu_i  \\
& +\Delta\omega_i + \tilde{\kappa}_H\left(\phi_A X_A+\phi_B X_B+\phi_S X_S\right)\Big\} \,dt \\ 
& + D_i \int_{t_n}^{t_{n+1}}  \Big\{ \Delta\omega_i 
+\tilde{\kappa}_H\left(\phi_A X_A+\phi_B X_B+\phi_S X_S\right)
 \Big\} \, dt
\end{split}
\end{aligned}
\label{eq: dynamische_gleichung_omega_integriert}
\end{equation}

Since the first integral on the right hand side of equation (\ref{eq:
dynamische_gleichung_omega_integriert}) contains the chemical potential minus
the leading stiffness contributions, we approximate it by an explicit Euler
formula, i.e.
\begin{equation}
\begin{aligned}
\begin{split}
& \int_{t_n}^{t_{n+1}}\Big\{
\Delta\mu_i
+\Delta\omega_i + \tilde{\kappa}_H\left(\phi_A X_A+\phi_B X_B+\phi_S X_S\right)\Big\} \,dt = \\ 
& h\Big\{ 
\Delta\mu_i^{(n)}
+\Delta\omega_i^{(n)} + \tilde{\kappa}_H\left(\phi_A^{(n)} X_A^{(n)}+\phi_B^{(n)} X_B^{(n)}+\phi_S^{(n)} X_S^{(n)}\right) \Big\} \\
& \eqqcolon h E_i^{(n)}.
\end{split}
\end{aligned}
\label{eq: Expliziter_Term_Abkuerzung}
\end{equation}

The second integral on the right hand side of equation (\ref{eq:
dynamische_gleichung_omega_integriert}) is treated semi-implicitly. To
approximate the included integrals $\int \phi_i X_i\, dt$ we use the quadrature
formulas
\begin{equation}
\int_{t_n}^{t_{n+1}} \phi_i \Delta\omega_i \, dt \approx \phi^{(n)}\Delta\omega_i^{(n+1)} h
\label{eq7: imex_integration_phi_laplace}
\end{equation}
and 
\begin{equation}
\int_{t_n}^{t_{n+1}} \phi_i\nabla\omega_i\cdot \nabla\omega_i \, dt \approx
\phi_i^{(n)}\nabla\omega_i^{(n)}\cdot \nabla\omega_i^{(n+1)}h,
\label{eq7: imex_integration_gradientquadrat}
\end{equation}
which are first order time accurate. The first $\nabla\omega_i$ on the right
hand side of equation (\ref{eq7: imex_integration_gradientquadrat}) is treated
explicitly to obtain a linear system of equations for $\omega_i^{(n+1)}$.
Inserting the short hand notations from equations (\ref{eq: X_S_abkuerzung}),
(\ref{eq: X_AB_abkuerzung}), and (\ref{eq: Expliziter_Term_Abkuerzung})
together with the quadrature formulas (\ref{eq7: imex_integration_phi_laplace})
and (\ref{eq7: imex_integration_gradientquadrat}) into equation (\ref{eq:
dynamische_gleichung_omega_integriert}) yields the semi-implicit integration
scheme shown in figure \ref{fig: semiimpliziter_Integrator}, equations
(\ref{eq7: neues_schema_omega_A}), (\ref{eq7: neues_schema_omega_B}), and
(\ref{eq7: neues_schema_omega_S})).

Spatial derivatives are discretized by a second order finite differences. To
solve the resulting discrete linear system of equations for
$\omega_i^{(n+1)}(\vec{r})$ we use the Generalized Minimal Residual Method
(GMRES), which is a Krylow subspace iteration method for linear systems with
positive semidefinite matrices and known to be efficient and robust
\cite{Saad.1986}.  In our particular implementation a Krylow iteration is
considered to have converged once the residual drops below $10^{-10}$ or the
maximum number of 50 iterative steps is reached. In case the residual is still
above $10^{-8}$ after 50 steps, we decrease the width of subsequent time steps
by multiplication with $1/1.5$. Further details on the implementation
of the algorithm can be found in \cite{simon_diss}.  We set the damping
coefficients to $\alpha=0.5$ and $\beta_A=\beta_B=\beta_S=0$. For this choice,
the simulations in the present work are found to be sufficiently stable. 

\section*{REFERENCES}
\bibliography{Literaturverzeichnis_Veroeffentlichung2}

\begin{thebibliography}{10}

\bibitem{Herlach.2010}
D.~M. Herlach, I.~Klassen, P.~Wette, and D.~Holland-Moritz.
\newblock {Colloids as model systems for metals and alloys: A case study of
  crystallization}.
\newblock {\em Journal of Physics: Condensed Matter}, 22(15):153101, 2010.

\bibitem{Smallenburg.2011}
F.~Smallenburg, N.~Boon, M.~Kater, M.~Dijkstra, and R.~Roij.
\newblock {Phase Diagrams of Colloidal Spheres with a Constant Zeta-Potential}.
\newblock {\em Journal of Chemical Physics}, 134:074505, 2011.

\bibitem{Bucaro.2009}
M.~A. Bucaro, P.~R. Kolodner, J.~A. Taylor, A.~Sidorenko, J.~Aizenberg, and
  T.~N. Krupenkin.
\newblock {Tunable Liquid Optics: Electrowetting-Controlled Liquid Mirrors
  Based on Self-Assembled Janus Tiles}.
\newblock {\em Langmuir}, 25(6):3876--3879, 2009.

\bibitem{Lensen.2008}
D.~Lensen, D.~M. Vriezema, and {van Hest, J. C. M.}
\newblock {Polymeric Microcapsules for Synthetic Applications}.
\newblock {\em Macromolecular Bioscience}, 8(11):991--1005, 2008.

\bibitem{Discher.2002}
D.~E. Discher and A.~Eisenberg.
\newblock {Polymer Vesicles}.
\newblock {\em Science}, 297:967--972, 2002.

\bibitem{Zhang.2008}
L.~Zhang, F.~X. Gu, J.~Chan, A.~Z. Wang, R.~S. Langer, and O.~Farokhzad.
\newblock {Nanoparticles} in {Medicine}: {Therapeutic} {Applications} and
  {Developments}.
\newblock {\em Clinical Pharmacology and Therapeutics}, 83:761--769, 2008.

\bibitem{Gindy.2009}
M.~E. Gindy and R.~K. Prud'homme.
\newblock Multifunctional nanoparticles for imaging, delivery and targeting in
  cancer therapy.
\newblock {\em Expert Opinion on Drug Delivery}, 6(8):865--878, 2009.

\bibitem{Vartholomeos.2011}
P.~Vartholomeos, M.~Fruchard, A.~Ferreira, and C.~Mavroidis.
\newblock {MRI-Guided Nanorobotic Systems for Therapeutic and Diagnostic
  Applications}.
\newblock {\em Annual Review of Biomedical Engineering}, 13(1):157--184, 2011.

\bibitem{Allen.2004}
T.~M. Allen and P.~R. Cullis.
\newblock {Drug} {Delivery} {Systems}: {Entering} {the} {Mainstream}.
\newblock {\em Science}, 303:1818--1822, 2004.

\bibitem{Bleul.2014}
R.~Bleul.
\newblock {\em {Herstellung, Characterisierung und Funktionalisierung polymerer
  Nanopartikel und Untersuchung der Wechselwirkung mit biologischen Systemen}}.
\newblock Dissertation, Freie Universität Berlin, 2014.

\bibitem{Discher.1999}
B.~M. Discher, Y.~Y. Won, D.~S. Ege, J.~C.~M. Lee, F.~S. Bates, D.~E. Discher,
  and D.~A. Hammer.
\newblock {Polymersomes: Tough Vesicles Made from Diblock Copolymers}.
\newblock {\em Science}, 284:1143--1146, 1999.

\bibitem{Thiermann.2014}
R.~Thiermann.
\newblock {Selbstorganisation amphiphiler Block-Copolymere in Mikromischern}.
\newblock {\em Dissertation, Berlin University of Technology}, 2014.

\bibitem{Thiermann.2012}
R.~Thiermann, W.~M{\"u}ller, A.~Montesinos-Castellanos, D.~Metzke, P.~L{\"o}b,
  V.~Hessel, and M.~Maskos.
\newblock Size controlled polymersomes by continuous self-assembly in
  micromixers.
\newblock {\em Polymer}, 53(11):2205--2210, 2012.

\bibitem{Tenzer.2011}
S.~Tenzer, D.~Docter, S.~Rosfa, A.~Wlodarski, J.~Kuharev, A.~Rekik, S.~K.
  Knauer, C.~Bantz, T.~Nawroth, C.~Bier, J.~Sirirattanapan, W.~Mann, L.~Treuel,
  R.~Zellner, M.~Maskos, H.~Schild, and R.~H. Stauber.
\newblock {Nanoparticle Size Is a Critical Physicochemical Determinant of the
  Human Blood Plasma Corona: A Comprehensive Quantitative Proteomic Analysis}.
\newblock {\em ACS Nano}, 5(9):7155--7167, 2011.

\bibitem{Maeda.2000}
H.~Maeda, J.~Wu, T.~Sawa, Y.~Matsumura, and K.~Hori.
\newblock Tumor vascular permeability and the epr~effect in macromolecular
  therapeutics: a review.
\newblock {\em Journal of Controlled Release}, 65:271--284, 2000.

\bibitem{Zhang.2012}
C.~Zhang, V.~J. Pansare, R~.K. Prud'homme, and R.~D. Priestley.
\newblock Flash nanoprecipitation of polysterene nanoparticles.
\newblock {\em Soft Matter}, 8:86--93, 2012.

\bibitem{Mueller.2009}
W.~M{\"u}ller.
\newblock {Hydrophobe und hydrophile Beladung polymerer Vesikel}.
\newblock {\em Dissertation, Johannes Gutenberg University Mainz}, 2009.

\bibitem{Nikoubashman.2016}
A.~Nikoubashman, V.~E. Lee, C.~Sosa, R.~K. Prud'homme, R.~D. Priestley, and
  A.~Z. Panagiotopoulos.
\newblock {Directed Assembly of Soft Colloids through Rapid Solvent Exchange}.
\newblock {\em ACS Nano}, 10:1425--1433, 2016.

\bibitem{Falk.2010}
L.~Falk and J.~M. Commenge.
\newblock Performance comparison of micromixers.
\newblock {\em Chemical Engineering Science}, 65(1):405--411, 2010.

\bibitem{Drese.2004}
K.~S. Drese.
\newblock {Optimization of interdigital micromixers via analytical
  modeling---exemplified with the SuperFocus mixer}.
\newblock {\em Chemical Engineering Journal}, 101(1-3):403--407, 2004.

\bibitem{Schonfeld.2004.2}
F.~Sch{\"o}nfeld, K.~S. Drese, S.~Hardt, V.~Hessel, and C.~Hofman.
\newblock {Optimized distributive} $\mu$-mixing {by 'chaotic' multilamination}.
\newblock {\em NSTI-Nanotech}, 1:378--381, 2004.

\bibitem{simon_diss}
S.~Ke{\ss}ler.
\newblock Dissertation Universit\"at Mainz, 2017.
\newblock in preparation.

\bibitem{meinArtikel1}
S.~Ke{\ss}ler, F.~Schmid, and K.~Drese.
\newblock {Modeling size controlled nanoparticle precipitation with the
  co-solvency method by spinodal decomposition}.
\newblock {\em Soft Matter}, 12:7231 -- 7240, 2016.

\bibitem{Wall.1966}
M.~E. Wall, M.~C. Wani, C.~E. Cook, K.~H. Palmer, McPhail~A. T., and G.~A. Sim.
\newblock {Plant Antitumor Agents. I. The Isolation and Structure of
  Camptothecin, a Novel Alkaloidal Leukemia and Tumor Inhibitor from
  Camptotheca acuminata}.
\newblock {\em Journal of American Chemical Society}, 88:3888 -- 3890, 1966.

\bibitem{Torchilin.2005}
V.~P. Torchilin.
\newblock {Recent advances with liposomes as pharmaceutical carriers}.
\newblock {\em Nature Review Drug Discovery}, 4:145 -- 160, 2005.

\bibitem{WMueller.2011}
W.~M{\"u}ller, K.~Koynov, S.~Pierrat, R.~Thiermann, C.~Fischer, and M.~Maskos.
\newblock {pH-change protective PB-b-PEO polymersomes}.
\newblock {\em Polymer}, 52:1263 -- 1267, 2011.

\bibitem{Kabanov.2002}
A.~V. Kabanov, E.~V. Batrakova, and V.~Y. Alakhov.
\newblock {Pluronic((R)) block copolymers for overcoming drug resistance in
  cancer}.
\newblock {\em Advanced Drug Delivery Reviews}, 54:759 -- 779, 2002.

\bibitem{Maurits.1997b}
N.~M. Maurits and J.~G. E.~M. Fraaije.
\newblock {Mesoscopic dynamics of copolymer melts: From density dynamics to
  external potential dynamics using nonlocal kinetic coupling}.
\newblock {\em The Journal of Chemical Physics}, 107(15):5879, 1997.

\bibitem{He.2006}
X.~He and F.~Schmid.
\newblock {Dynamics of Spontaneous Vesicle Formation in Dilute Solutions of
  Amphiphilic Diblock Copolymers}.
\newblock {\em Macromolecules}, 39(7):2654--2662, 2006.

\bibitem{He.2008}
X.~He and F.~Schmid.
\newblock {Spontaneous Formation of Complex Micelles from a Homogeneous
  Solution}.
\newblock {\em Physical Review Letters}, 100(13), 2008.

\bibitem{Muller.2005}
M.~M{\"u}ller and F.~Schmid.
\newblock {Incorporating Fluctuations and Dynamics in Self-Consistent Field
  Theories for Polymer Blends}.
\newblock In {\em {Advances in Polymer Science}}, volume 185, pages 1--85.
  Springer Verlag, 2005.

\bibitem{Edwards.1965}
S.~F. Edwards.
\newblock {The statistical mechanics of polymers with excluded volume}.
\newblock {\em Proceedings of the Physical Society}, 85:613 -- 624, 1965.

\bibitem{Schmid.1998}
F.~Schmid.
\newblock Self-consistent-field theories for complex fluids.
\newblock {\em Journal of Physics: Condensed Matter}, 10:8105--8138, 1998.

\bibitem{Helfand.1975}
E.~Helfand.
\newblock {Theory of inhomogeneous polymers: Fundamentals of the Gaussian
  random-walk model}.
\newblock {\em The Journal of Chemical Physics}, 62(3):999, 1975.

\bibitem{Helfand.1972}
E.~Helfand and Y.~Tagami.
\newblock {Theory of the Interface between Immiscible Polymers. II}.
\newblock {\em The Journal of Chemical Physics}, 56(7):3592, 1972.

\bibitem{Fredrickson.2006}
G.~H. Fredrickson.
\newblock {\em The {Equilibrium Theory} of {Inhomogeneous Polymers}}.
\newblock Oxford University Press, 2006.

\bibitem{Kawasaki.1987}
K.~Kawasaki and K.~Sekimoto.
\newblock {Dynamical Theory of Polymer Melt Morphology}.
\newblock {\em Physica}, 143A:349--413, 1987.

\bibitem{Uneyama.2007}
T.~Uneyama.
\newblock Density functional simulation of spontaneous formation of vesicle in
  block copolymer solutions.
\newblock {\em The Journal of Chemical Physics}, 126(11):114920, 2007.

\bibitem{weiss1}
T.~M. Weiss, T.~Narayanan, and M.~Gradzielski.
\newblock {Dynamics of spontaneous vesicle formation in fluorocarbon and
  hydrocarbon surfactant mixtures}.
\newblock {\em {Langmuir}}, {24}({8}):{3759--3766}, {APR 15} {2008}.

\bibitem{adams1}
D.~J. Adams, S.~Adams, D.~Atkins, M.~F. Butler, and S.~Furzeland.
\newblock {Impact of mechanism of formation on encapsulation in block copolymer
  vesicles}.
\newblock {\em {J. Contr. Release}}, {128}({2}):{165--170}, {2008}.

\bibitem{gummel1}
J.~Gummel, M.~Sztucki, T.~Narayanan, and M.~Gradzielski.
\newblock {Concentration dependent pathways in spontaneous self-assembly of
  unilamellar vesicles}.
\newblock {\em {Soft matter}}, {7}({12}):{5731--5738}, {2011}.

\bibitem{Sofonea.1999}
V.~Sofonea and K.~R. Mecke.
\newblock Morphological characterization of spinodal decomposition kinetics.
\newblock {\em The European Physical Journal B}, 8(1):99--112, 1999.

\bibitem{Nielsen.2014}
T.~T. Nielsen.
\newblock {\em {An Implementation Of The Connected Component Labelling
  Algorithm}}.
\newblock
  {https://www.codeproject.com/articles/825200/an-implementation-of-the-connec%
ted-component-label}, 2014.

\bibitem{Mantz.2008}
H.~Mantz, K.~Jacobs, and K.~Mecke.
\newblock {Utilizing Minkowski functionals for image analysis: A marching
  square algorithm}.
\newblock {\em Journal of Statistical Mechanics: Theory and Experiment},
  2008(12):P12015, 2008.

\bibitem{JohannesArtikel}
J.~Heuser, G.~J.~A. Sevink, and F.~Schmid.
\newblock Self-assembly of polymeric particles in poiseuille flow: A hybrid
  lattice boltzmann/external potential dynamics simulation study.
\newblock {\em Macromolecules}, 50(11):4474--4490, 2017.

\bibitem{Franeker.2015}
J.~J. van Franeker, G.~H.~L. Heintges, C.~Schaefer, G.~Portale, W.~Li, M.~M.
  Wienk, P.~van~der Schoot, and R.~A.~J. Janssen.
\newblock {Polymer Solar Cells: Solubility Controls Fiber Network Formation}.
\newblock {\em Journal of the American Chemical Society}, 137:11783--11794,
  2015.

\bibitem{Freed.1995}
K.~F. Freed.
\newblock Interrelation between density functional and self-consistent-field
  formulations for inhomogeneous polymer systems.
\newblock {\em The Journal of Chemical Physics}, 103(8):3230, 1995.

\bibitem{Dahlquist.1963}
G.~Dahlquist.
\newblock A special stability problem for linear multistep methods.
\newblock {\em Communications of the ACM}, 14(3):176--179, 1963.

\bibitem{Zhu.1999}
J.~Zhu, L.~Q. Shen, J.~Shen, and V.~Tikare.
\newblock {Coarsening kinetics from a variable-mobility Cahn-Hilliard equation:
  Application o fa semi-implicit Fourier spectral method}.
\newblock {\em Physical Review E}, 60(4):3564--3572, 1999.

\bibitem{Ascher.1995}
U.~M. Ascher, S.~J. Ruuth, and B.~T.~R. Wetton.
\newblock Implicit-explicit methods for time-dependent partial differential
  equations.
\newblock {\em SIAM Journal on Numerical Analysis}, 32(3):797--823, 1995.

\bibitem{Tzeremes.2002}
G.~Tzeremes, K.~O. Rasmussen, T.~Lookman, and A.~Saxena.
\newblock {Efficient computation of the structural phase behavior of block
  copolymers}.
\newblock {\em Physical Review E}, 65:041806 1--5, 2002.

\bibitem{Saad.1986}
Y.~Saad and M.~H. Schlutz.
\newblock {GMRES}: {A} generalized minimal residual algorithm for solving
  nonsymmetric linear systems.
\newblock {\em SIAM Journal on Scientific and Statistical Computing},
  7:856--869, 1986.

\end{thebibliography}

\end{document}